\documentclass[twocolumn]{jpsj2} 
\title{A Possible Phase Transition in $\beta$-pyrochlore Compounds}

\author{Kazumasa HATTORI and Hirokazu TSUNETSUGU}

\inst{Institute for Solid State Physics, University of Tokyo, 5-1-5, Kashiwanoha, Kashiwa Chiba 277-8581, Japan
}

\abst{We investigate a lattice of interacting anharmonic oscillators by using
 a mean field theory and exact diagonalization.
We construct an effective five-state hopping model with intersite repulsions
 as a model for $\beta$-pyrochlore AOs$_2$O$_6$(A=K, Rb or Cs).
We obtain the first order phase transition line 
from large to small oscillation amplitude phases as temperature decreases.
 We also discuss the possibility of a phase with local electric polarizations.
 Our theory can explain the origin of the mysterious first order transition in KOs$_2$O$_6$.
 }

\kword{$\beta$-pyrochlore, anharmonic phonon\vspace{-2cm}}

\begin{document}
\maketitle
\vspace{-0.5cm}
Recently, various unusual low-temperature properties of 
 low-energy anharmonic phonons have attracted much 
attention. In $\beta$-pyrochlore compounds AOs$_2$O$_6$ (A=K, Rb or Cs),\cite{HiroiUnprecedented, Hiroi2ndAno,SummaryHiroi} low-energy anharmonic oscillations of A-ions are 
considered to play an important role for the realization of the unusual electrical resistivity\cite{HiroiUnprecedented, Hiroi2ndAno}
 and specific heat\cite{SummaryHiroi}, the strong 
coupling superconductivity,
\cite{NMR_RbOs2O6_2005,BattloggSC,IsotropicGappSCSC}
 and the nuclear magnetic relaxation time,\cite{Yoshida} 
especially in KOs$_2$O$_6$. 
In these compounds, A-ions are located in large Os$_{12}$O$_{18}$ cages and form a 
diamond lattice structure. The cage opens in the four [111] 
directions, which are the bond directions of the diamond lattice,
 and it is expected that A-ion oscillations are anharmonic and have 
large amplitudes along these four 
directions.\cite{Yamaura,NeutronSasai} Since the atomic radius of K is the 
smallest among A-ions, the effects of the anharmonicity
 is most prominent in KOs$_2$O$_6$. 

One of the most interesting phenomena in
 KOs$_2$O$_6$ is the first order transition below the 
superconducting transition temperature $T_c=9.6$ K.
\cite{Hiroi2ndAno}
 The transition temperature
 $T_p=$7.5 K is robust upon applying magnetic fields up to 14 T.\cite{SummaryHiroi} 
This suggests that 
the transition is driven by lattice degrees of freedom. Interestingly, 
it is found that, in the non-superconducting state under magnetic fields, 
the temperature $(T)$ dependence of resistivity 
 changes from an unusual $\sqrt{T}$ behavior to a $T^2$ dependence 
expected in the fermi liquid  theory.\cite{SummaryHiroi} Corresponding to 
this, the life time of quasiparticles 
becomes longer below $T_p$.\cite{ThermalCondKasahara,peneShimono}
The energy gap of the superconductivity observed in the 
photoemission spectroscopy
 also shows an anomaly at $T_p$.\cite{PES} 
These experimental results suggest that quasiparticles are coupled to 
the lattice degrees of freedom leading to the first order transition.

As for the theories, 
it is proposed from the band structure calculation, 
that the effective potential of A-ion 
is shallow and even has a minimum at off-center positions, 
especially in KOs$_2$O$_6$.\cite{BandCal}
Dahm and Ueda investigated the origin of the unusual temperature 
dependence of the resistivity and the saturation behavior of the 
NMR relaxation rate at high temperatures.\cite{Dahm} 
 As for the first order transition, Kune\v{s} 
{\it et al}., proposed that it is multiple-q 
ordering of the potassium displacements.\cite{FrustKunes}
 This scenario, however, contradicts the
observations that  symmetry 
is not broken below $T_p$ as various experiments show.
\cite{SummaryHiroi,Yoshida,Raman}

In this Letter, we extend the model used in Ref. 15 to include 
quantum hoppings by introducing five localized states of the ionic configurations. One of the five states is 
an on-center state and others are off-center ones. 
We analyze this model by using a 
mean field theory and exact diagonalization (ED) calculations.
In our theory, the first order transition is characterized by
 the A-ion density at the on-center position, which is 
 an order parameter in a sense of a liquid-gas transition.
 We will discuss implications of the present results 
for KOs$_2$O$_6$ and physical consequences of this phase
 transition.

%
%
Before going into detailed calculations, we discuss the first order transition
 without symmetry breaking in KOs$_2$O$_6$ from a 
phenomenological point of view.
 The local 
symmetry around A-ion is point group $T_d$ and local 
potential $V({\bf r})$ of A-ion is approximated as
$
V({\bf r})=
a |{\bf r}|^2+bxyz+c|{\bf r}|^4 
+c'(x^4+y^4+z^4)+O(|{\bf r}|^5)
$,
 where 
${\bf r}=(x,y,z)$ is the coordinates of A-ion from the on-center 
position  of the diamond lattice. It is noted that the third order term $bxyz$ 
can exist. Due to this term, $V(\bf{r})$ becomes shallow 
in the four [111] directions of the bonds in the diamond lattice. 
This makes s-wave $(\varphi_s)$ and f-wave 
($\varphi_{xyz}$) functions of 
oscillation states hybridized. 
Both of them belong to $A_1$ representation 
in $T_d$ point group. 
The one-body ground state $\Psi$ for the potential $V({\bf r})$ is 
 a linear combination of 
$\varphi_s$, $\varphi_{xyz}$ and other $A_1$ components: 
$\Psi=c_s\varphi_s+c_{xyz}\varphi_{xyz}+\cdots$.
 The point is that $\Psi$ has 
inner degrees of freedom characterized by the coefficients 
$c_s$ and $c_{xyz}$, {\it etc.}
These coefficients can vary without breaking any symmetry. 
In this Letter, we propose a minimal model to exhibit such variation 
as functions of $T$ and flatness of $V(\bf r)$. 
When this variation becomes a sudden jump, the 
first order transition occurs. 
In this scenario, the $b$-term plays a more important 
role for anharmonicity and 
anisotropy than the fourth order terms.
It is noted that this kind of transition is 
completely different in nature from classical structural
 transitions, which
 are described by the instability of some phonon modes.

\begin{figure}[tb]
	\begin{center}
    \includegraphics[width=0.44\textwidth]{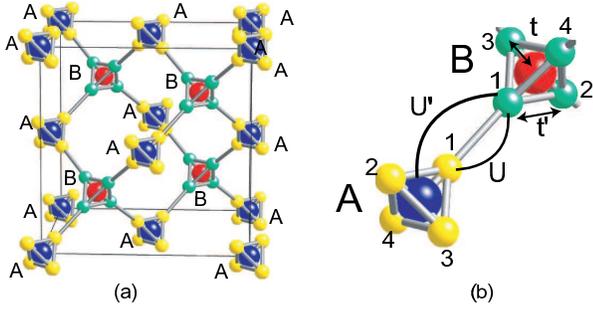}
\end{center}
\vspace{-0.3cm}
\caption{(a) Cubic unit cell of diamond lattice with one on-center and four off-center sites. On- and off-center positions are indicated by large and small filled circles, respectively. (b) Interactions and tunneling matrix elements.}
\label{fig-1}
\vspace{-0.3cm}
\end{figure}

Now, we introduce a minimal model to describe the phase transition at
 $T_p$ in KOs$_2$O$_6$. We investigate a Hamiltonian of five 
localized states with inter-site repulsive interactions 
which are originated from Coulomb interactions between A$^+$ ions.
 The five localized states are
one on-center $(\xi=0)$ and four off-center $(\xi=1,2,3$ 
and $4)$ states as shown in Fig. \ref{fig-1}.
These five states can describe two kinds of $A_1$ components: 
$\varphi_s$ and $\varphi_{xyz}$
 and also p-wave like components $\varphi_{p_i}(i=x,y$ and $z)$  
which have the same symmetry $T_2$ as one-phonon states.
 The on-center state is located exactly at the diamond lattice 
site ${\bf x}_{{\bf i}\alpha}
=(x_{{\bf i}\alpha},y_{{\bf i}\alpha},z_{{\bf i}\alpha})$ 
where ${\bf i}$ belonging to sublattice 
$\alpha(=$ $A$ or $\bar{A}\equiv B$). 
Others are located at positions 
${\bf x}_{{\bf i}\alpha}+{\bf d}_{\alpha \xi}$ 
shifted 
along the bond directions of the diamond lattice. 
We can regard that the parameter $|{\bf d}_{\alpha \xi}|$ 
is implicitly included in parameters in our model Hamiltonian shown below.  
We note that $b$-term in $V({\bf r})$ prefers wavefunctions having 
large weight in the bond directions, and thus Coulomb interactions for 
these wavefunctions become effectively large.
These bases rather than phonon ones are expected 
to be a good starting point to discuss AOs$_2$O$_6$, 
since the local oscillations of A-ion is strongly anharmonic and 
large in the directions of 
the bonds especially for A=K. This choice of the bases corresponds to write $\Psi$ 
in the form of $c_a\varphi_a + c_{a'}\varphi_{a'}$. Here, $\varphi_a$ and 
$\varphi_{a'}$ correspond to the on-center state and a symmetric 
linear combination of off-center ones, respectively. 

\begin{table}[!b]
\vspace{-0.5cm}
   \begin{tabular}{lcc}
         symmetry & energy & wavefunction\\
         \hline
         \hline
       $A_1^{(1)}$ & $\epsilon_{A_1^{(1)}}\equiv e-\delta$ & $c_a\varphi_a + c_{a'}\varphi_{a'}$ \\
       $A_1^{(2)}$ & $\epsilon_{A_1^{(2)}}\equiv e+\delta$ & $c_{a'}\varphi_a - c_{a}\varphi_{a'}$ \\
       $T_2$ & $\epsilon_{T_2}\equiv \bar{\epsilon}^{\alpha}+t'$ & $\{ \varphi_{x},\varphi_{y},\varphi_{z}\}$\\ 
\hline
   \end{tabular}
\vspace{-0.3cm}
\caption{Eigenstates and eigenenergies of the mean field Hamiltonian. 
$e\equiv\frac{1}{2}(\bar{\epsilon}^{\alpha}+\bar{\epsilon}_{00}^{\alpha}-3t')$, 
$\delta\equiv 
 \frac{1}{2}\sqrt{(\bar{\epsilon}_{00}^{\alpha}-\bar{\epsilon}^{\alpha}+3t')^2+16t^2}$,
$c_a\equiv2t/\sqrt{(e-\delta-\bar{\epsilon}^{\alpha}_{00})^2+4t^2}$ and 
$c_{a'}\equiv(\bar{\epsilon}^{\alpha}_{00}-e+\delta)/\sqrt{(e-\delta-\bar{\epsilon}^{\alpha}_{00})^2+4t^2}$. The wavefunctions are written as 
      $\varphi_a\equiv\phi_0$,
      $\varphi_{a'}\equiv\frac{1}{2}(\phi_1+\phi_2+\phi_3+\phi_4)$,
      $\varphi_x \equiv \frac{1}{2}(\phi_1-\phi_2+\phi_3-\phi_4)$,
      $\varphi_y \equiv \frac{1}{2}(\phi_1-\phi_2-\phi_3+\phi_4)$, and
      $\varphi_z \equiv \frac{1}{2}(\phi_1+\phi_2-\phi_3-\phi_4)$,
 where we omit the site indices.
}
\label{tbl-1}
\end{table}

The Hamiltonian in this Letter is written as
\begin{eqnarray}
H=\sum_{{\bf i}\alpha \xi\eta}\epsilon^{\alpha}_{\xi\eta}
\phi^{\dagger}_{{\bf i}\alpha \xi} \phi_{{\bf i}\alpha \eta}
+\sum_{{\bf ij}\alpha\beta \xi\eta}U_{{\bf ij}\xi\eta}^{\alpha\beta}
n_{{\bf i}\alpha \xi}n_{{\bf j}\beta \eta},\label{eq1}
\end{eqnarray}
where $\phi^{\dagger}_{{\bf i}\alpha \xi}$ represents a creation 
operator of the ionic state $|{{\bf i}\alpha \xi}\rangle$ localized 
at position $\xi$ at a site $\bf i$ of sublattice $\alpha$ and 
$
n_{{\bf i}\alpha \xi}=\phi^{\dagger}_{{\bf i}\alpha \xi}\phi_{{\bf i}\alpha \xi}
$. 
 The five states labeled by $\xi$ and $\eta$ have 
hopping matrix elements $\epsilon^{\alpha}_{\xi\eta}$ with $\xi\ne \eta$ which represent
 the local quantum  kinetic energy. We set 
$\epsilon_{0\xi}^{\alpha}\equiv -t $ for $\xi\ge 1$ 
and $\epsilon_{\xi\eta}^{\alpha}\equiv-t'$ for $\xi,\eta\ge 1$. 
For the local energy levels, we set 
$\epsilon_{00}^{\alpha}=0$ and 
$\epsilon_{\xi\xi}^{\alpha}\equiv \epsilon $ for $\xi\ge 1$. 
For the interaction
terms, we use only the nearest neighbor Coulomb repulsion $U$ and the 
next nearest one $0<U'<U$ as shown in Fig. \ref{fig-1} (b). 
This is because the screening by the electrons on the cage is not effective for these two terms because of the peculiar structure of the cage, but effective for others, and thus we neglect them.

\begin{figure}[!t]
	\begin{center}
    \includegraphics[width=0.38\textwidth]{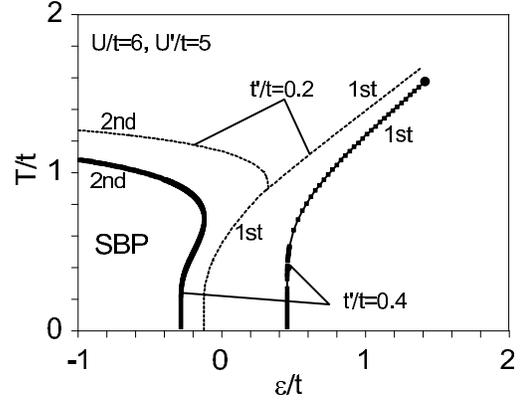}
\end{center}
\vspace{-0.5cm}
\caption{Mean field $\epsilon$-$T$ phase diagram of model (\ref{eq1}). 
$U=6t$, $U'=5t$ and $t'/t=0.4$ and $0.2$.}
\label{fig-2}
\vspace{-0.3cm}
\end{figure}

First, we investigate this model by a mean field approximation (MFA) in the
 case that the translational symmetry is not broken.
The mean field decoupling of the interactions leads 
 the effective local energies 
$\bar{\epsilon}^{\alpha}_{00}=4U'\langle n_{\bar{\alpha}}\rangle$
 and 
$
\bar{\epsilon}^{\alpha}=\epsilon+U\langle n_{\bar{\alpha}}\rangle+U'\langle 
n_{\bar{\alpha}0}\rangle
$, if the local symmetry $T_d$ is not broken. 
Here, $\langle \cdot \rangle$ denotes the thermal average and 
$\langle n_{\alpha} \rangle=\langle n_{{\bf i}\alpha \xi}\rangle$
with $\xi\ge 1$ and 
$\langle n_{\alpha 0} \rangle=\langle n_{{\bf i}\alpha 0}\rangle$. 
Now we can easily 
diagonalize the mean-field Hamiltonian, and show
 the eigenenergies 
and the eigenstates in 
Table \ref{tbl-1}. There are two singlet ($A_1$) and one 
triplet ($T_2$) states. The ground state $A_1^{(1)}$ is 
an in-phase combination of $\varphi_a$ and $\varphi_{a'}$ but the excited $A_1^{(2)}$ state is an out-of-phase one. 
Using these eigenstates, we calculate 
$\langle n_{\alpha 0} \rangle=1-4\langle n_{\alpha} \rangle$ 
such that the self-consistency is satisfied.

 When we consider a symmetric phase with 
$\langle n_{A0}\rangle=\langle n_{B0}\rangle\equiv n_0$ at $T=0$, the self-consistency condition reads 
\begin{eqnarray}
2f(n_0)=\beta^{-1}(n_0)-\beta(n_0),\label{self}
\end{eqnarray}
 with $\beta(n_0)=\sqrt{(1-n_0)/n_0}$ and $f(n_0)=[\epsilon+U/4-U'-3t'+(2U'-U/4)n_0]/4t$.
 A first order transition at $T=0$ occurs if
$
U'-U/8>4t
$ as $\epsilon$ is varied, since 
this condition is satisfied, Eq. (\ref{self}) has three solutions. 
By using Eq. (\ref{self}), 
the ground state energy per site $E_g$ is represented 
by the function of $n_0$ as 
$
E_g=(U'-U/8)(1-n_0)^2-2t\beta(n_0)
$.

In Fig. \ref{fig-2}, we show the mean field $\epsilon$-$T$ 
phase diagram for $U/t=0.6$, $U'=0.5t$ and $t'/t=0.4$ and $0.2$. 
There are the first order transition
 line, and the second order 
transition to 
a symmetry broken phase (SBP). 
Between the two transitions,
 there is a region where 
$n_0$ remains $\sim 1/5$ down to $T=0$ for $t'/t=0.4$. 
In this region, the oscillation of A-ion 
becomes large and gains the
 kinetic energy rather 
than the potential energy. 
This intermediate region becomes smaller
 as $t'$ decreases and eventually 
the FE phase touches the first order transition line.
When we increases $U$, the former is suppressed as
 understood from the $T=0$ analysis above 
and becomes a crossover. On the other hand, the transition 
temperature to SBP is enhanced. This is because only $U$ enters 
in the self-consistent equation for SBP, and the other parameters are
 implicitly included in it through the energy levels.

First, we discuss the first order 
transition which is
 relevant to $\beta$-pyrochlore compounds, 
and then investigate the properties of SBP. 
Note that we consider the case of $\epsilon < 0$ is not realized in 
AOs$_2$O$_6$.

\begin{figure}[t]
	\begin{center}
    \includegraphics[width=0.5\textwidth]{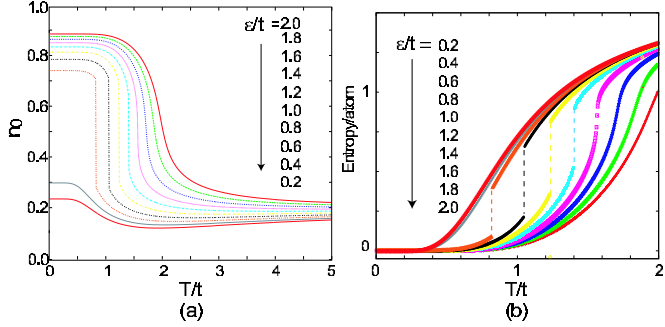}
\end{center}
\vspace{-0.5cm}
\caption{Temperature dependence of (a) $n_0$ and (b) entropy. 
$U=6t$, $U'=5t$ and $t'/t=0.4$.}
\label{fig-3}
\vspace{-0.3cm}
\end{figure}

In Fig. \ref{fig-3} (a), the temperature dependence of $n_0$ is 
shown for various values of $\epsilon$.
At high temperatures, the density 
$n_0$ is about 1/5. This means that the five positions are equally 
populated due to thermal fluctuations. As for the energy level,
for $\epsilon=0.8t$, 
$\epsilon_{T_2}-\epsilon_{A_1^{(1)}}\sim 3t$ and 
$\epsilon_{A_1^{(2)}}-\epsilon_{A_1^{(1)}}\sim 5t$ 
in the high temperature limit.
In the case of 
large $\epsilon$, there is no discontinuous change but a smooth 
crossover to a large $n_0$ value with decreasing temperature. 
As $\epsilon$ decreases, the crossover
 becomes a jump, {\it i.e.,} the first order transition.
Just below the first order transition temperature $T^*$, $n_0$ suddenly 
increases in order to reduce the potential 
energy $\epsilon$ and the intersite repulsions. 
Consequently, $\epsilon_{T_2}$ {\it increases} below $T^*$. 
This means that the effective potential around the central position 
becomes steeper below $T^*$. This is understood by considering the 
$T=0$ case. When $U'-U/8>0$, which is automatically satisfied 
by that for the first order transition at $T=0$, $\epsilon_{T_2}$ 
increases  as $n_0$ increases. 

In Fig. \ref{fig-3} (b), we show the temperature 
dependence of the entropy.
The entropy at high temperature approaches to $\log5\simeq 1.6$. 
There is a jump of entropy for $0.6\le\epsilon/t\le1.4$, 
accompanying the first order transition.
The released entropy at $T^*$ is 
$\sim\frac{1}{2}\log2$ per A-ion for $\epsilon/t=1.0$. The jump of entropy 
becomes small as $\epsilon$ increases from $\epsilon/t=1.0$ and 
disappears at the critical end point 
$\epsilon\sim 1.4t$ and $T\sim 1.6t$ and also at $T^*\to 0$.

In the SBP, the average  
$
\langle n_{{\bf i}A1}+n_{{\bf i}A2}-n_{{\bf i}A3}-n_{{\bf i}A4} \rangle
=\langle n_{{\bf i}B3}+n_{{\bf i}B4}-n_{{\bf i}B1}-n_{{\bf i}B2} \rangle
$ 
becomes finite uniformly. This means 
a finite polarization ${\bf P}\parallel$ [001], {\it i.e.} ferro-electric (FE) state. This phase is stabilized when $T_2$ states have 
a low excitation energy. This corresponds to
small and/or negative $\epsilon$. The reentrant behavior 
as a function of temperature
is due to the fact that the FE fluctuations arise from the 
excited $T_2$ states. 
It is noted that the FE state is six-fold degenerate (${\bf P}\parallel \pm[100],\ \pm [010]$ and $\pm[001]$). For the three-dimensional six-state Potts model,
 it is known that the transition is the first order both from 
MFA and Monte Carlo simulations.\cite{Blote} Our results of MFA is different from the result of MFA for the Potts model. This is because the symmetry ${\bf P}\leftrightarrow-{\bf P}$ is present in this model.


We have examined all the possible uniform symmetry broken 
patterns in our MFA. 
The phase with uniform polarizations
is the most stable one. 
For example, a phase where the $B$-sublattice polarization 
is antiparallel to the $A$-sublattice one is not realized, 
because of repulsive interactions. 
The energy of another 
FE phase with ${\bf P}\parallel$ [111] is slightly higher than that
 of ${\bf P}\parallel$ [001]. When we consider an order with 
the wavevector ${\bf Q}_x=(2\pi,0,0)$, the order parameter is 
$
\langle n_{{\bf i}A1}+n_{{\bf i}A2}-n_{{\bf i}A3}-n_{{\bf i}A4} \rangle
=\langle n_{{\bf i}B2}+n_{{\bf i}B3}-n_{{\bf i}B1}-n_{{\bf i}B4} \rangle
\ne 0$. This corresponds to polarization ${\bf P}_A\parallel$ [001] and 
${\bf P}_B\parallel$ [010] in the two sublattices. This state is one 
of the twelve-fold degenerate states: 
${\bf P}_\alpha\to -{\bf P}_\alpha$, $A\leftrightarrow B$ 
and ${\bf Q}_x \to {\bf Q}_y$ or ${\bf Q}_z$.
 The energy of these states become lower than that of 
the FE state when long-range repulsions are included.


\begin{figure}[b]
\vspace{-0.3cm}
	\begin{center}
    \includegraphics[width=0.5\textwidth]{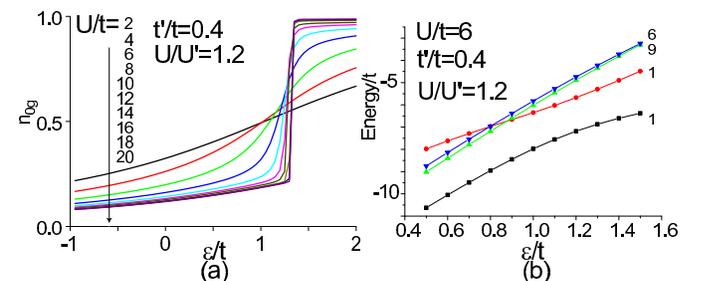}
\end{center}
\vspace{-0.5cm}
\caption{(a) Occupation number $n_{0g}$ as 
functions of $U$ and $\epsilon$. (b) Energy eigenvalues of four low-energy multiplets. Numbers represent the degeneracy. $U/U'=1.2$ and $t'/t=0.4$.}
\label{fig-4}
\vspace{-0.3cm}
\end{figure}

Next, in order to investigate the effect 
of fluctuations beyond the MFA, we carry out 
 ED calculations of this model. 
We use the cubic unit cell 
which includes eight A-ions and retains the $T_d$ 
symmetry with periodic boundary conditions. 
Since there is no conserving quantities such 
as spin or charge as in 
electron systems, the ED
in larger system size is difficult at present. 
However, we can see the effects of the 
fluctuations in this size.

The expectation value
$ n_{{\bf i}A0} \equiv n_{0g}$ 
in the ground state is shown 
in Fig. \ref{fig-4} (a) as functions 
of $U$ and $\epsilon$. Here, $n_{0g}$ is 
independent on the site indices. 
We find a sharp change of 
$n_{0g}$ for relatively 
larger $U$ than in the case of mean field results. 
This may be due to the effects of the quantum 
fluctuations and finite size. 
In this calculation, the sharp change 
of $n_{0g}(\epsilon)$ is 
due to the anti-crossing of the ground and the 
first excited state as shown in Fig. \ref{fig-4} (b). 
Since the off-diagonal elements
 of the Hamiltonian (\ref{eq1}) are all non-positive, 
we can apply the Perron-Frobenius theorem and
 prove that the ground state is always 
unique. This means there is no level crossing
 for the ground state.
We expect this anti-crossing becomes narrower with 
increasing the system size, and 
evolves into a first order transition in the infinite size limit.

The presence of SBP is also reflected in the structure of 
excited states. When $\epsilon$ is smaller than $\sim t$, 
the first excited states are nine-fold degenerate as shown in 
Fig. \ref{fig-4} (b).
As discussed before, six ${\bf Q}={\bf 0}$ states and 
twelve ${\bf Q}\ne{\bf 0}$ states are degenerate in the MFA.
In this finite size cluster, these eighteen states are coupled
 and split into a few multiplets and the lowest excited multiplet 
has nine states. This unusual large degeneracy is due to
 the special topology of eight-ion
 cluster. Each $A$-sublattice site is
 connected to all $B$-sublattice sites, and vise versa, and 
the total symmetry turns out to be $S_4^A \otimes S_4^B \otimes Z_{2}$, 
where $S_4^{\alpha}$ and $Z_{2}$ represent $S_4$ group for 
sublattice $\alpha$ and the permutation $A\leftrightarrow B$,
 respectively. This 
special symmetry is broken in larger system sizes and it 
remains as an interesting question how the degeneracy is lifted.

The tendency of the first 
order transition from small to large $n_{0g}$ as $\epsilon$ increases,
is consistent with the results of MFA 
and due to the quantum fluctuations, the transition occurs 
at larger $U$ and $U'$ than in the MFA.
In order to conclude whether the intermediate region 
obtained in the MFA, {\it i.e.}, 
the region $\epsilon\sim 0$ for $t'/t=0.4$ in Fig. \ref{fig-2}
 exists or not,
it is needed to carry out detailed analysis
 of the finite size scaling for the energy of excited states.



%
%

As we discussed above, this model shows the first order 
transition at $T^*$ below which the on-center 
density suddenly increases. 
We believe that this corresponds to 
the first order transition at $T_p$ in KOs$_2$O$_6$. 
Since anharmonicity for Rb and Cs compounds is weaker 
than that for K, we can regard that the cases for 
Rb and Cs correspond to 
the larger $\epsilon$ region in the phase diagram where there is no 
transition but a smooth crossover. The fact that this 
transition does not break any 
symmetries is consistent with experimental 
results.\cite{SummaryHiroi,Yoshida,Raman} 
Recently, Yamaura {\it et al.}, observed that the 
charge distributions of potassium shrink below $T_p$ by the neutron 
experiments in KOs$_2$O$_6$.\cite{yamauraprivate} This corresponds to
the increase of $n_0$. 
 In addition to this, the suppression of the 
quasiparticle scattering below $T_p$\cite{SummaryHiroi,ThermalCondKasahara,peneShimono} is explained as a result of the sudden freezing of the A-ion oscillations. It is also noted that 
$T^*$ is typically one third of $\epsilon_{T_2}$ for 
$\epsilon=0.8t$ in Fig. \ref{fig-2}. This is 
qualitatively in good agreement with estimated Einstein energy 
$\sim 22$ K\cite{SummaryHiroi} and $T_p=7.5$ K.

In real materials, the cage 
is composed of O and Os ions.
Dynamics of these ions will be affected by the transition 
at $T_p$. In $\beta$-pyrochlore structure, O$^{2-}$ ions 
can move without breaking lattice symmetry. Their positions 
are expected to be different in high- and 
low-temperature phases. When we take into account the effect of 
the nearest neighbor
 K$^+$-ion from an O$^{2-}$-ion, the O$^{2-}$ 
is expected to move toward the K$^{+}$-ion below $T_p$ due to the increase of the attractive Coulomb energy. 

An interesting prediction for KOs$_2$O$_6$ of our theory 
is that the $T_p$ increases when we apply 
 pressure. This 
is understood by noting that the energy level of off-center states $\epsilon$ 
is expected to increase, since $\epsilon$ corresponds to the 
steepness of the potential $V(\bf r)$. 
It is interesting to explore the critical end 
point of the first order transition at high pressure. The critical exponents of this transition are expected to be those of 
the three dimensional Ising model.
Interestingly, It is observed that the pressure dependence of 
the superconducting transition temperature $T_c$ has a small 
anomaly at around 2 GPa and $T$=8 K.\cite{Miyoshi} This might be 
a signature of enhanced $T_p$. 

A recent theoretical study of the strong coupling 
superconductivity for KOs$_2$O$_6$ pre-assumes the 
energy of the low-energy Einstein phonon decreases 
below $T_p$.\cite{Thalmeier} This assumption is not 
consistent with our results, since the energy level 
of the $T_2$-states increases below $T^*$. The temperature 
dependence of the energy gap of the superconductivity 
across $T_p$\cite{PES} might be explained by the 
balance between the increase of the phonon energy and 
the reduction of coupling constants with conduction electrons.

 Theoretically, it is important to extend the present 
model to that can describe the higher energy physics, 
including the couplings with conduction electrons\cite{HatUD} 
and acoustic phonons. The validity of using the 
short range interaction should also be explained by a 
microscopic theory, since if one uses the bare Coulomb or 
a simple Yukawa type interaction, the first order transition 
does not occur in the MFA discussed in this Letter. 
Understanding the first order transition in the phonon basis 
and the observed unusual $T^5$ specific heat 
in KOs$_2$O$_6$\cite{SummaryHiroi} still remains as a big 
challenge and needs further theoretical studies. 

In conclusion, we have investigated the origin of the 
first order transition in KOs$_2$O$_6$. We have 
proposed that the transition at $T_p$ in KOs$_2$O$_6$ 
is an on- and off-center transition without symmetry breaking 
which is liquid-gas type. 
Our results show that such a transition is possible 
in the case of rather shallow potentials. 
Since the potential in KOs$_2$O$_6$ is expected to be 
the shallowest in three compounds of AOs$_2$O$_6$ 
(A=K, Rb and Cs), this explains why only KOs$_2$O$_6$
 shows the first order transition.

\section*{Acknowledgment}
The authors would thank Y. Tomita and Z. Hiroi for grateful discussions. This work is supported by KAKENHI (No. 19052003 and No. 20740189).

\end{document}